# Deep learning based high-resolution incoherent x-ray imaging with a single-pixel detector

Yu-Hang He[1,2]†, Ai-Xin Zhang[1,2]†, Ming-Fei Li[3,7], Yi-Yi Huang[1,2], Bao-Gang Quan[1,2], Da-Zhang Li[5], Ling-An Wu[1,2]* and Li-Ming Chen[1,2,4,6]*

[1]*Institute of Physics, Chinese Academy of Sciences, Beijing 100191, China*
[2]*University of Chinese Academy of Sciences, Beijing 100049, China*
[3]*Beijing Institute of Aerospace Control Devices, Beijing 100039, China*
[4]*Songshan Lake Material Laboratory, Guangdong 523808, China*
[5]*Institute of High Energy Physics, Chinese Academy of Sciences, Beijing 100049, China*
[6] *IFSA Collaborative Innovation Center*, *Shanghai Jiao Tong University, Shanghai 200240, China*
[7]*Quantum Engineering Research Center, China Aerospace Science and Technology Corporation, Beijing 100094, China*

X-ray "ghost" imaging has drawn great attention for its potential to lower radiation dose in medical diagnosis. For practical implementation, however, the efficiency and image quality have to be greatly improved. Here we demonstrate a computational ghost imaging scheme where a bucket detector and specially designed modulation masks are used, together with a new robust deep learning algorithm in which a compressed set of Hadamard matrices is incorporated into a multi-level wavelet convolutional neural network. By this means we have obtained an image of a real object from only 18.75% of the Nyquist sampling rate, using a portable tabletop incoherent x-ray source of ~37 μm diameter. A high imaging resolution of ~10 μm is achieved, which represents a concrete step towards the realization of a practical low cost x-ray ghost imaging camera for applications in biomedicine, archeology, material science, and so forth.

Safety and image quality are the two major factors in x-ray imaging. In traditional schemes, according to the Ross criterion[1] there has to be a balance between radiation dose and image quality since high resolution and good contrast require a sufficiently long exposure, which means greater dose. Brilliant phase contrast images with nanometer resolution can be obtained in state-of-the-art synchrotron facilities that provide monochromatic ultra-bright x-ray beams[2,3]. For most users, however, incoherent x-ray sources for use in a laboratory are more accessible, but in this case the resolution is limited by the source size[4,5]. So how to increase the resolution and lower the dose and cost of x-ray imaging with such polychromatic sources is a significant problem.

Different from traditional imaging, ghost imaging (GI) is a second-order correlation based technology which retrieves information about an object from a series of reference patterns and the corresponding intensity values measured by just a single-pixel ("bucket") detector[6]. As this type of detector is more sensitive than an array of pixels, even in poor weather[7] or ultra-low exposure situations[8] GI can still retrieve an acceptable image. The key point is to generate or record a series of reference speckle patterns that illuminate the object plane, then by convoluting the total intensity measured at the bucket detector with the reference patterns the image can be retrieved. Since the speckle patterns are similar to the matrix arrays in pixel detectors, the resolution is limited by the average size of the speckles. However, through various means GI can now achieve resolutions beyond the Rayleigh criterion[9], and has already been applied to many fields such as microscopy[10,11], lidar detection, and remote sensing[12–14].

---

† Contributed equally to this work as co-first authors.
* Corresponding author. Email: wula@iphy.ac.cn (L.A.W.); lmchen@iphy.ac.cn (L.M.C.).

Over the past few decades, GI has been demonstrated with quantum light[6], classical light[15] and even particles—atoms[16] and electrons[17]. There is great potential in the fabrication of cheap high resolution GI cameras at terahertz[18] and infrared[10] wavelengths that cannot use silicon sensors. Ghost tomography has broad prospects in various fields[19–21]. At visible wavelengths it is easy to obtain reference patterns with the aid of a beamsplitter, which can furthermore be replaced by a spatial light modulator and the number of light paths reduced to only one as in computational ghost imaging (CGI)[22,23]. At x-ray wavelengths, however, a major difficulty is that there are no suitable optics. Several approaches have been proposed to overcome this problem. In the beam splitter strategy[24–27], a crystal can be taken as a beamsplitter based on Laue diffraction if the flux density of the source in a narrow bandwidth is very strong; even so, mechanical vibration may blur the image if not controlled strictly. Another strategy is pre-recording[28,29]; with a phase or amplitude modulation plate a series of repeatable speckle patterns can be generated quite easily, but saving and transferring the large amount of data required entails much extra time, which greatly reduces practicability. Another problem is that all the existing schemes rely on a high resolution x-ray camera for calibration, which increases the cost, while the resolution is limited by the pixel size of the CCD (or CMOS) arrays. Recently, x-ray GI of a one-dimensional slit has been realized with a single-pixel detector but with the bright monocromatic beam from a synchrotron source[30]. For common practical applications such as in biomedicine, a resolution of several microns must be achieved to be of real use in diagnosis.

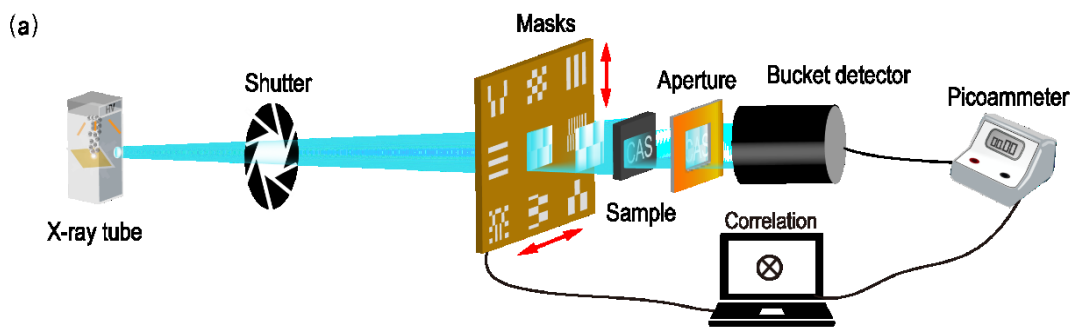

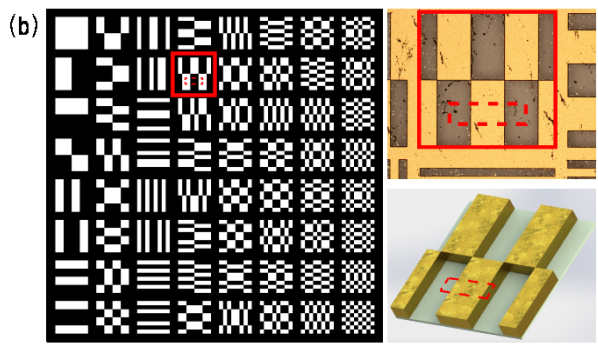

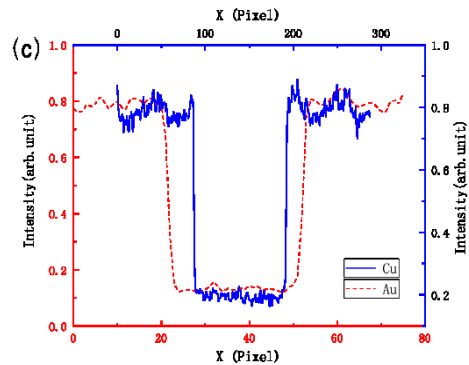

**Fig. 1.** (Color online) (a) Experimental setup of CXGI. The set of masks is mounted on a motorized 2D translation stage. (b) Part of the specially designed masks, with a typical pattern (enclosed in red) shown enlarged in the upper right, and its 3D visualization in the lower right. (c) Cross section of the portion enclosed by the red dashed line in (b); blue solid line: Cu on a laminate substrate, with a modulation depth ratio of 75%; red dashed line: Au on a $SiO_2$ substrate, with a modulation ratio of 83%.

In this letter, we report a computational x-ray GI (CXGI) scheme by which high resolution images were obtained with only an inexpensive single-pixel detector and a portable incoherent x-ray source, plus the use of a deep learning algorithm. Instead of our previous pre-recording scheme, to dispense with the array detector we used a motorized stage to project a set of computer-designed modulation masks onto the object. The layout of the experiment is shown in Fig. 1(a). As shown, the incoherent hard x-ray beam from a micro-focus x-ray tube (Incoatec Source Iμs) first passes through an adjustable shutter and then through a certain matrix in the modulation mask, which is mounted on a 2-dimension motor stage. The modulated x-ray pattern illuminates the sample, is partially transmitted, and its intensity measured by a bucket detector. Another aperture blocks out unwanted light from around the object. After each exposure the shutter is closed, the mask translated to the next adjacent matrix, and the measurement repeated. The single-pixel detector used in our experiment was an x-ray diode (Hamamatsu) with a beryllium window and wrapped in a copper shell, which converted the incident intensity to a current of several pico-amperes. Of course, a CCD camera could be used as a bucket detector by integrating the total intensity registered on all the pixels, but this requires a much longer processing time and is much more expensive, while a true bucket detector has better sensitivity. After several measurements the image is retrieved by an appropriate algorithm.

The amplitude modulation board was fabricated from a metal layer etched into a series of patterns upon a flat substrate, the former being strongly absorbing and the latter transparent to hard x-rays. Two different masks were made: the first was an inexpensive printed circuit board composed of a 100 μm thick copper foil on a 500 μm thick laminate substrate; this was used to test the feasibility of our scheme. The second consisted of a 10 μm thick layer of gold foil electroplated onto a 4 inch square, 500 μm thick $SiO_2$ substrate; this board was made for high resolution CXGI. Both masks were etched with a set of Hadamard matrices[31]; the number and size of the pixels of the Cu and Au masks were 32×32, 150 μm and 64×64, 10 μm, respectively. A detailed description of the matrix design is given in the Supplementary Material: "Pattern selection". An illustration of part of the masks is shown in Fig. 1(b). An enlargement of the area outlined by the red solid line is shown in the upper right, and the lower right is the corresponding 3-dimension (3D) visualization. We define the modulation depth ratio $D_r$ as $D_r = (D_{max} - D_{min})/ D_{max}$, where $D_{max}$ and $D_{min}$ represent the maximum and minimum intensities recorded by the bucket detector, respectively. The cross-section of the part enclosed by the red dashed line in Fig. 1(b) is shown in Fig. 1(c), from which we see that the modulation depth ratio $D_r$ of the Cu and Au masks is about 75% and 83%, respectively. This profile was plotted from the gray values of a direct image on a CCD camera of the x-ray transmission through the mask.

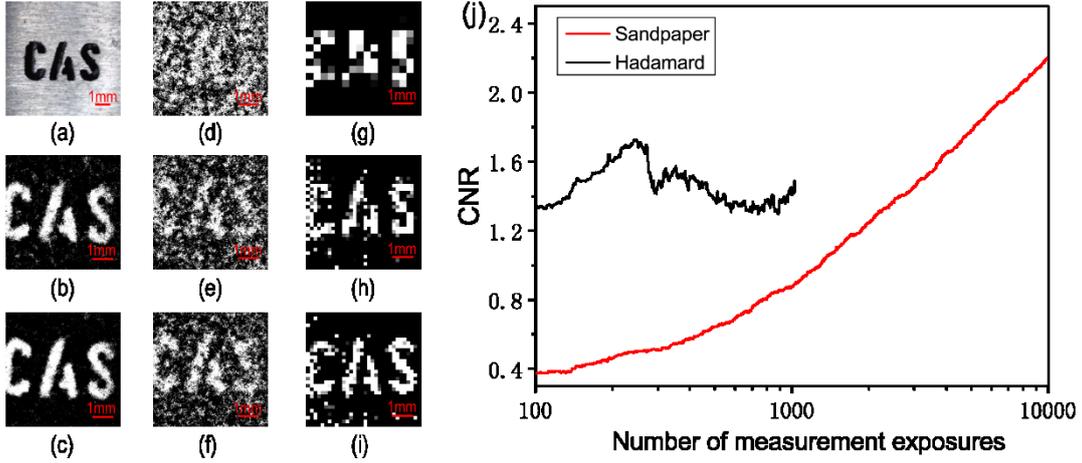

**Fig. 2.** (Color online) (a) Object: 5 mm thick stainless steel mask with stenciled letters "CAS". (b)–(f): sandpaper XGI images retrieved after 5000, 10000, 128, 512 and 1024 exposures. (g)–(i): CXGI images retrieved with the Cu mask after 128, 512 and 1024 exposures. (j) CNR vs. number of exposures for XGI (red line) and CXGI (black line).

When the Cu mask was used, a 5 mm thick stainless steel mask with stenciled letters "CAS" was chosen as the object, as shown in Fig. 2(a). We compared the performance of two different modulation means: random sandpaper speckles and a set of pre-designed Hadamard matrices. The same second-order correlation algorithm was taken for fairness. By adjusting the magnification and speckle size, the resolution was set at 150 μm for both cases. We adopt the contrast-to-noise ratio (CNR) as a criterion of image quality, defined as

$$CNR \equiv \frac{\langle G_1 \rangle - \langle G_0 \rangle}{\sqrt{\sigma_1^2 + \sigma_0^2}}, \tag{1}$$

where $G_1$ and $G_0$ are the GI values for any pixel where the transmission is 1 or 0, respectively; $\sigma_1^2$ and $\sigma_0^2$ are the corresponding variances, i.e. $\sigma_1^2 = \langle G_1^2 \rangle - \langle G_1 \rangle^2$. When sandpaper is used, since the modulation is random and its speckles are quite uniformly distributed, it can perform well when the number of exposures equals or exceeds the number of pixels. This can be seen from Figs. 2 (b) and (c), which correspond to 5000 and 10000 exposures, respectively; the retrieved image contains about 3500 pixels in this case. Figures 2(d)–(f) and (g)–(i) are the results of the sandpaper speckles and Hadamard masks for 128, 512, and 1024 measurements, respectively. We see that the Hadamard mask performs much better than sandpaper under the same number of measurements. Figure 2(j) provides a more quantitative comparison, where the CNR of both methods is plotted as a function of the number of exposures. Here, for the same CNR, when the Hadamard board is used, the number of measurements is reduced by an order of magnitude due to the orthogonality of the mask patterns. However, the CNR begins to decrease when the exposure number exceeds 300. In addition, it is evident from Fig. 2(i) that even in the full Nyquist sampling case there is still some unwanted noise. There are two explanations for this: one is imperfections of the Hadamard patterns due to uneven etching during the electrochemical processing, which become more pronounced in the finer patterns containing more complex structures; the other is that the finer structures produce smaller fluctuations of the x-ray intensity which therefore cannot be easily detected when the detector is not sensitive enough. The first problem could be solved with more precise lithography, and the second by either increasing the source intensity or improving the sensitivity of the detector, e.g. by using a photomultiplier tube.

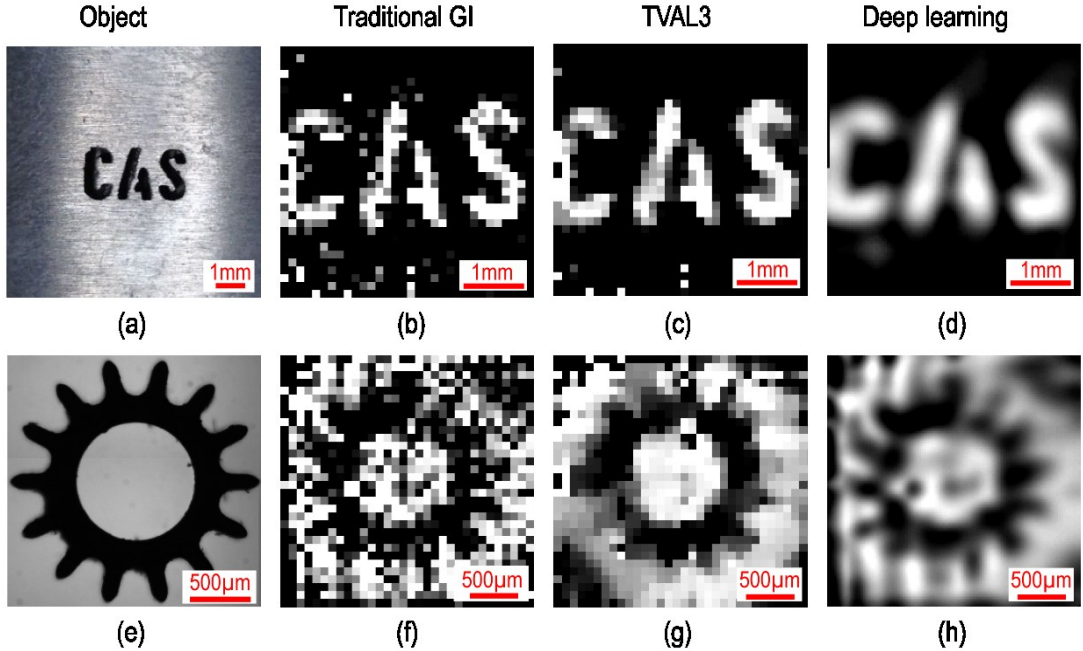

**Fig. 3.** (Color online) Results of XGI with the Cu modulation mask and their corresponding images. (a) Same object as in Fig. 2(a). (b) – (d) Images retrieved by traditional GI, TVAL3, and deep learning, after 1024 exposures, with CNR values of 1.49, 1.75 and 2.43, respectively. (e) Object: a metal gear with 14 teeth. (f) – (h): Images of metal gear retrieved by traditional GI, TVAL3 and deep learning after 1024 exposures, with CNR values of 0.6, 0.5 and 1.56, respectively.

Aside from the updates in hardware device, the CXGI image can also be further improved by specially designed algorithms. Compressive sampling (CS) has been widely used in GI and single-pixel cameras[32–34]. For the same bucket signals of the "CAS" object mentioned above, Fig. 3(c) shows the image reconstructed by the total variation augmented Lagrangian alternating direction algorithm (TVAL3), one of the most popular CS algorithms[35]. The image is greatly improved as it has less noise and the edges are much sharper than in Fig. 3(b). The CNR values of Figs. 3(b) and (c) are 1.49 and 1.75, respectively. To test the universality of this algorithm, we take a more complex sample as the object which is a metal gear with 14 teeth, about 1.5 mm in diameter. The CNR of Fig. 3(g) retrieved by TVAL3 is 0.5, which is even a little bit poorer than the 0.6 of Fig. 3(f) obtained by traditional GI. It seems that the CS algorithm is not robust enough when the image contains complex structure and the Hadamard mask is imperfect.

Deep neural networks are computational models which learn representations of data with multiple levels of abstraction[36]. They are proving very successful at discovering features in high-dimensional data in many areas including GI. Recently, deep learning was successfully used to recover structured signals (in particular, images) from their under-sampled random linear measurements[37–39]. With an irregular trained basis, it can process an image at high speed even under a 2% exposure compression ratio[40]. However, the rectangular Hadamard matrices that we chose (because these shapes are relatively easy to fabricate) are not suitable for the deep convolutional auto-encoder network used in the deep learning algorithm mentioned above. Thus we developed a new compressible Hadamard plus multi-level wavelet convolutional neural network (CH-MWCNN) algorithm, which takes the receptive field size and computational efficiency into consideration and can be rather robust even with an imperfect modulation mask, as in our case. With this we succeeded in obtaining greatly improved images, as shown in Figs. 3(d) and (h) where the CNR ratios are 2.43 and 1.56, respectively, higher than all the other methods. Admittedly, there is still some blurring present due to the fact that the size of the modulation mask is only 32×32 pixels, which means that the sparser pixels cannot bring out enough detail while the finer structures tend to be over-fitted by our CH-MWCNN algorithm. This could be greatly improved by increasing the density of the pixels, though of course this would require better precision in the mask fabrication.

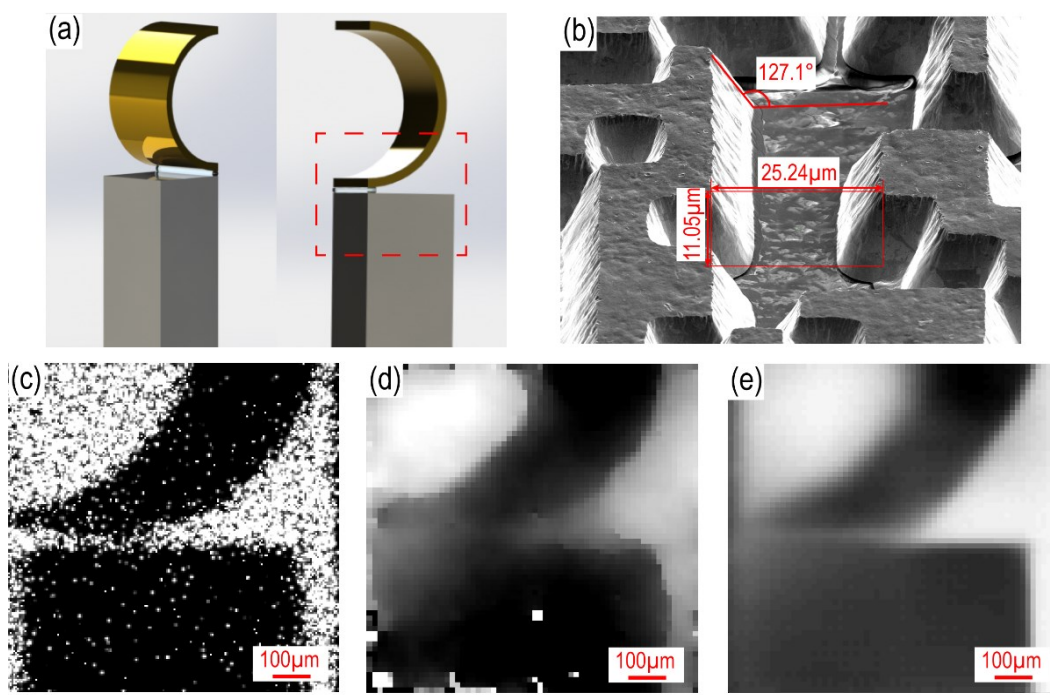

**Fig. 4.** (Color online) Object and images with a gold modulation mask. (a) Two 3D views of the object from different angles where the area enclosed by the red dashed line is the actually exposed part of the object. (b) SEM image of part of the Au mask. (c) Direct x-ray absorption/projection image of the sample with 5 s exposure time. (d) CXGI image retrieved by TVAL3; its CNR is 0.27. (e) Image reconstructed by CH-MWCNN; its CNR is 2.65.

To improve both hardware and software, the Cu mask was replaced by a 64×64 Au mask, with pixels of size 10 μm square. A semi-cylindrically shaped object made of gold and glued to a rectangular column was used here as our object. Its 3D visualizations from different angles are shown in Fig. 4(a); the exposed area of the object was actually just 0.64×0.64 mm. A direct absorption/projection x-ray image of the object is presented in Fig. 4(c), where the exposure time was 5 sec. Figure 4(d) shows the image recovered by TVAL3; we can see it is seriously blurred and the CNR is only 0.27. Similar to the problems in Cu mask manufacture, there is also distortion in the ion etched Au mask, part of which is shown in the scanning electron microscope (SEM) image of Fig. 4(b), taken at an angle of 52°. Here we observe the sloping profile of the etched edges, which ideally should be perpendicular. This is probably the reason why the TVAL3 algorithm failed to give better detail. On the other hand, when CH-MWCNN was used, the CNR of the retrieved image, shown in Fig. 4(e), improved significantly; here it is 2.65. The gap between the semi-cylinder and the rectangular column which is ~10 μm thick can be distinguished clearly. This result was obtained under a sampling rate of 18.75% with 0.3 s recovery time, a performance much better than TVAL3 and certain other compressed sensing algorithms. The average processing time of CH-MWCNN was between 0.2—1 s for an image of 64×64 pixels, depending on its complexity, running on a laptop with an Intel® Core™ i7-6600U central processing unit and 12 GB random access memory.

The x-ray source size in our experiment was 30×37 μm, measured by a knife edge method. In traditional projection radiography, according to geometrical optics the image will be blurred when the object is smaller than half the source size; this is the key limit of resolution both in traditional absorption and phase contrast x-ray imaging. In contrast, due to its nonlocal feature, the resolution of CXGI depends on the design and quality of the illumination masks, which in our amplitude modulation scheme are quite good. The resolution of our current experiments is in fact limited by the mask lithography technology, and so far we have achieved a value of several microns, as can be seen from Fig. 4. Our low sampling rate means lower dose, less measurement time and faster processing, which are all essential for real applications. Of course the image quality needs to be improved further, but the real object images retrieved by our CH-MWCNN algorithm fully indicate the huge potential of our CXGI scheme. If applied to a certain field such as medical diagnosis where training can be acquired with the vast clinical image data resources available, the results should certainly be much better.

In conclusion, we have realized CXGI with an incoherent x-ray source and a true bucket detector, with both simulation and experimental results showing that we have surpassed the resolution limit of incoherent x-ray imaging even at subsampling rates. The setup is simple, cost effective and convenient to operate. Compared with random speckle modulation, the pre-designed masks have a consistent speckle size so that resolution can be predetermined, and combined with certain orthogonal matrices such as Hadamard matrices, the measurement efficiency can be greatly improved. A new CH-MWCNN algorithm has been implemented, by which even when the modulation mask contains some distortions we can still observe fine structures under a low subsampling rate, regardless of the complexity of the object. As a result, images with 10 μm resolution have been obtained for a source size of about 37 μm at a sampling rate of 18.75%, which indicates that the measurement time as well as the radiation dosage in x-ray diagnosis can both be greatly reduced. The resolution could be improved further if finer and better masks and more sensitive detectors were used, in addition to optimization of other technical aspects. The image quality would also be much better if imaging data for real samples could be used to strengthen our deep learning algorithm. Already, our current CXGI scheme demonstrates that it should be quite feasible to build a practical, low-cost, single-pixel x-ray camera for use in biomedicine, archeology and material diagnosis.

**Acknowledgements**

We thank Prof. Junjie Li for technical support during fabrication of the modulation masks. This work was supported by the National Key R&D Program of China (2017YFA0403301, 2017YFB0503301, 2018YFB0504302), National Natural Science Foundation of China (11334013, 11721404, U1530150, 11805266), the Key Program of CAS (XDB17030500, XDB16010200) and the Science Challenge Project (TZ2018005).

**Author contributions**

L. M. C. and L. A. W. proposed the research. Y. H. H. and A. X. Z. designed the key devices. Y. H. H., A. X. Z., B. G. Q. and D. Z. L. fabricated the key devices. Y. H. H., A. X. Z. and Y. Y. H. performed the experiments. A. X. Z. and M. F. L. developed the algorithm model. L. A. W., L. M. C., Y. H. H., A. X. Z. and M. F. L. contributed to the data analysis. Y. H. H., A. X. Z., L. A. W., M. F. L. and L. M. C. contributed to the writing of the manuscript.